\documentclass[conference]{IEEEtran}
\IEEEoverridecommandlockouts
\usepackage[nobreak]{cite}
\usepackage{amsmath,amssymb,amsfonts}
\usepackage{algorithmic}
\usepackage{graphicx}
\usepackage{textcomp}
\usepackage{subcaption}
\usepackage{xcolor}
\usepackage{booktabs}
\usepackage{url}
\usepackage{bm}
\usepackage{siunitx}
\usepackage{multirow}
\def\BibTeX{{\rm B\kern-.05em{\sc i\kern-.025em b}\kern-.08em
    T\kern-.1667em\lower.7ex\hbox{E}\kern-.125emX}}
\DeclareMathOperator*{\argmin}{argmin}

\newcommand{\txtrain}[0]{{}}
\newcommand{\txtarget}[0]{\prime}
\newcommand{\sinc}[0]{\text{sinc}}
\newcommand{\Cin}[0]{C^{\text{(in)}}}
\newcommand{\Cout}[0]{C^{\text{(out)}}}
\newcommand{\Kround}[0]{\check{K'}}

\begin{document}

\title{Algorithms of Sampling-Frequency-Independent Layers for Non-integer Strides}

\author{\IEEEauthorblockN{Kanami Imamura\dag,  Tomohiko Nakamura\dag, Norihiro Takamune\dag, Kohei Yatabe\ddag, and Hiroshi Saruwatari\dag}
\IEEEauthorblockA{\dag Graduate School of Information Science and Technology, The University of Tokyo, Tokyo, Japan}
\IEEEauthorblockA{\ddag Department of Electrical Engineering and Computer Science,\\ Tokyo University of Agriculture and Technology, Tokyo, Japan}
\thanks{This work was supported by JST, ACT-X Grant Number JPMJAX210G, Japan. T.~Nakamura is currently with the National Institute of Advanced Industrial Science and Technology (AIST).}}

\maketitle

\begin{abstract}
In this paper, we propose algorithms for handling non-integer strides in sampling-frequency-independent (SFI) convolutional and transposed convolutional layers.
The SFI layers have been developed for handling various sampling frequencies (SFs) by a single neural network.
They are replaceable with their non-SFI counterparts and can be introduced into various network architectures.
However, they could not handle some specific configurations when combined with non-SFI layers.
For example, an SFI extension of Conv-TasNet, a standard audio source separation model, cannot handle some pairs of trained and target SFs because the strides of the SFI layers become non-integers.
This problem cannot be solved by simple rounding or signal resampling, resulting in the significant performance degradation.
To overcome this problem, we propose algorithms for handling non-integer strides by using windowed sinc interpolation.
The proposed algorithms realize the continuous-time representations of features using the interpolation and enable us to sample instants with the desired stride.
Experimental results on music source separation showed that the proposed algorithms outperformed the rounding- and signal-resampling-based methods at SFs lower than the trained SF.
\end{abstract}


\section{Introduction} \label{sec:intro}
Deep neural networks (DNNs) have been used for various audio signal processing tasks such as music source separation \cite{mdx2021}, speech enhancement \cite{noise_challenge}, and automatic music transcription \cite{music-trans1,music-trans2}.
Most studies on DNN-based audio signal processing methods assume that the sampling frequency (SF) of an input signal is the same in the training and inference stages \cite{mdx2021,noise_challenge, music-trans1,music-trans2, convtasnet, unmix, spleeter2020, DenseNet, mrdla, hybrid-demucs, band-split-rnn, uss, dual-path-rnn, meta-learning, wavesplit, df-confromer}.
Hence, to handle untrained SFs, we need an additional processing such as signal resampling.

As an alternative to signal resampling, we previously proposed sampling-frequency-independent (SFI) convolutional layers \cite{sfi-convtasnet}.
These layers are based on the analogy between an ordinary convolutional layer and a collection of digital filters.
A digital filter can be designed from an analog filter.
Since an analog filter is SFI, we can use a collection of analog filters (latent analog filters) as an SFI structure.
By utilizing a digital filter design technique, we can consistently generate the parameters (weights) of the ordinary convolutional layer from the latent analog filters for various SFs.
We experimentally found that the SFI layers work more consistently than signal resampling for SFs much lower than a trained SF.
Since SF is usually task-specific, the SFI layers paved the way to realize an audio source separation method that can be universally used for any downstream tasks.

The SFI layers were applied to Conv-TasNet \cite{convtasnet}, which combines DNN-based mask predictors with a trainable analysis/synthesis filterbank.
This SFI extension is called SFI Conv-TasNet.
It uses the SFI layers only for the analysis/synthesis filterbank, leaving the mask predictors unchanged from Conv-TasNet, which enables us to introduce recently developed mask predictors \cite{uss, dual-path-rnn, meta-learning, wavesplit, df-confromer}.
It requires that for each target SF, the kernel size $K$ and stride $S$ of the SFI layer should be adjusted to match those used during training in seconds.
This is because the mask predictor is not SFI and its input should have the same time resolution for any SFs.

However, for some pairs of target and trained SFs, $K$ and $S$ become non-integers, even though a convolutional layer assumes them to be integers.
For example, when SFI Conv-TasNet is trained with $K=160$ and $S=80$ (5~\si{\milli\second} and 2.5~\si{\milli\second}, respectively) at an SF of 32~\si{\kilo\hertz}, the corresponding $K$ and $S$ are $110.25$ and $55.125$ at an SF of 22.05~\si{\kilo\hertz}, respectively.
Although the evaluation was performed at SFs where $K$ and $S$ became integers in \cite{sfi-convtasnet}, this problem is unavoidable to realize the versatile preprocessor of our interest.

One simple method to handle such a non-integer is to round it to the nearest integer.
However, rounding $S$ changes the time resolution of the mask predictor's input and degrades the separation performance, as we will show later in Section~\ref{sec:exp}.
Another method to handle non-integer strides is to use the rounding jointly with a mask predictor based on a two-dimensional convolutional neural network (2D-CNN).
Paulus \textit{et al.} experimentally showed that rounding $K$ and $S$ did not greatly affect the performance of speech dialogue separation when using a 2D-CNN-based mask predictor \cite{Paulus2022EUSIPCO}.
However, this method greatly restricts the architecture of the mask predictor.
Specifically, the mask predictor must be agnostic to time and frequency axes and must not contain any pooling or unpooling layers.
To ensure flexibility in the network architecture, we need to explore another approach.

In this paper, we propose an algorithm of the SFI convolutional layer for non-integer $S$ by utilizing windowed sinc interpolation.
This interpolation provides the continuous-time counterpart of a discrete-time signal.
Thus, by inserting it before the decimation in the convolutional layer, we can obtain the mask predictor's input having the desired stride.
We also extend this algorithm for an SFI version of a transposed convolutional layer (SFI transposed convolutional layer).
We demonstrate the effectiveness of the proposed algorithms in comparison to rounding- and signal-resampling-based methods through music source separation experiments.

\section{Conventional Methods}
\subsection{SFI Convolutional Layer}
We briefly describe a one-dimensional SFI convolutional layer \cite{sfi-convtasnet}.
Let $\Cin$ and $\Cout$ be the numbers of input and output channels of this layer, respectively.
This layer consists of $\Cin\Cout$ analog filters and an ordinary convolutional layer.
Given a target SF, it generates discrete-time impulse responses of length $K$ that are designed to approximate the continuous-time impulse response or frequency response of the analog filters by using time- or frequency-domain filter design methods, respectively.
Since the focus of this paper is how to handle non-integer $S$, we omitted the computation of the filter design methods due to space limitations (see \cite{sfi-convtasnet} for details).
A convolutional layer computes a cross-correlation between an input feature and the weights.
Hence, the discrete-time impulse responses are reversed in time and are used as the weights for the convolutional layer.
Replacing the convolutional layer with the transposed convolutional layer yields the SFI transposed convolutional layer.

\subsection{SFI Conv-TasNet}
\label{sec:sfi-convtasnet}
\begin{figure}
    \centering
    \includegraphics[width=\linewidth]{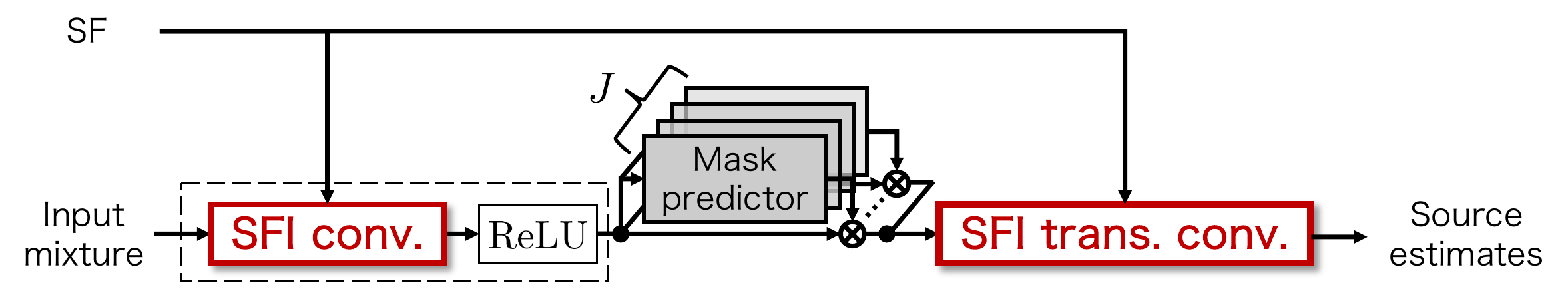}
    \caption{
    Architecture of SFI Conv-TasNet.
    ``SFI conv.'' and ``SFI trans. conv.'' denote SFI convolutional and transposed convolutional layers, respectively.
    }
    \label{fig:sfi-convtasnet}
\end{figure}

SFI Conv-TasNet is a Conv-TasNet extension that combines $J$ mask predictors and an \emph{SFI} analysis/synthesis filterbank \cite{sfi-convtasnet}, where $J$ is the number of sources.
Figure~\ref{fig:sfi-convtasnet} shows the architecture of SFI Conv-TasNet.
The analysis filterbank, called the encoder, consists of an SFI convolutional layer with $\Cin=1$ and $\Cout=C$ followed by a rectified linear unit (ReLU).
Given an $N$-length monaural mixture $\{x[n]\}_{n=0}^{N-1}$, the encoder outputs a pseudo time-frequency representation $\{X_{c}[m]\}_{c=0,m=0}^{C-1,M-1}$, where $M$ is the number of frames and $n$, $m$, and $c$ are the discrete-time, frame, and channel indices, respectively.
The pseudo time-frequency representation is fed into the mask predictor of source $j$, which mainly consist of one-dimensional dilated convolutional layers.
See \cite{meta-learning} for the details of the architecture of the mask predictors.
After multiplying $\{X_{c}[m]\}_{c,m}$ by the predicted mask, we obtain a time-domain separated signal of each source $\{\hat{s}_{j}[n]\}_{n=0}^{N-1}$ by using the synthesis filterbank, called the decoder.
The decoder is an SFI transposed convolutional layer with $C$ input channels and one output channel.
The kernel size $K$ and stride $S$ of the decoder are the same as those of the encoder.

At the inference stage, we need to adjust $K$ and $S$ in accordance with a target SF, as described in Section~\ref{sec:intro}.
Let $T$ be the sampling period of an input signal.
To clarify the values used in the training and inference stages, we hereafter use $K$, $S$, and $T$ for the training stage and $K^\txtarget$, $S^\txtarget$, and $T^\txtarget$ for the inference stage, respectively.
The kernel size and stride at $T^\txtarget$ are given as
\begin{equation}
    K^{\txtarget}=\dfrac{T^\txtrain}{T^\txtarget}K^{\txtrain},\quad S^{\txtarget}=\dfrac{T^\txtrain}{T^\txtarget}S^{\txtrain}. \label{eq:rounding}
\end{equation}
This adjustment keeps the time resolution of $\{X_{c}[m]\}_{c,m}$ unchanged, although it is valid only for integers $K^{\txtarget}$ and $S^{\txtarget}$.

\subsection{Windowed Sinc Interpolation}
The windowed sinc interpolation is a popular bandlimited interpolation of a discrete-time signal.
Let $t$ be continuous-time and $g(t)$ be the real-valued window function.
For a given discrete-time signal $x[n]$ with a sampling period of $T$, the interpolated signal $\tilde{x}(t)$ is given as
\begin{equation}
    \tilde{x}(t) = \sum_{n=-\infty}^{\infty} x[n] h(t-nT,T), \label{eq:sinc-interp}
\end{equation}
where $h(t,T)$ is the windowed sinc function:
\begin{equation}
    h(t,T)=g(t)\sinc\left( \dfrac{t}{T} \right), \quad \sinc(t)=\dfrac{\sin(\pi t)}{\pi t}.
\end{equation}
Now, we choose a window function with a finite support around $t=0$.
More concretely, we assume that $g(t)=0$ for $t<-LT/2$ or $t>LT/2$, where $L$ is a positive integer.
This choice can reduce the infinite sum in \eqref{eq:sinc-interp} to the finite sum from $n=\lceil t/T-L/2 \rceil$ to $n=\lfloor t/T+L/2\rfloor$, which enables us to implement the interpolation with computers.

\section{Proposed Method} \label{sec:proposed}
\begin{figure*}
    \centering
    \begin{minipage}{0.48\hsize}
    \includegraphics[width=\linewidth]{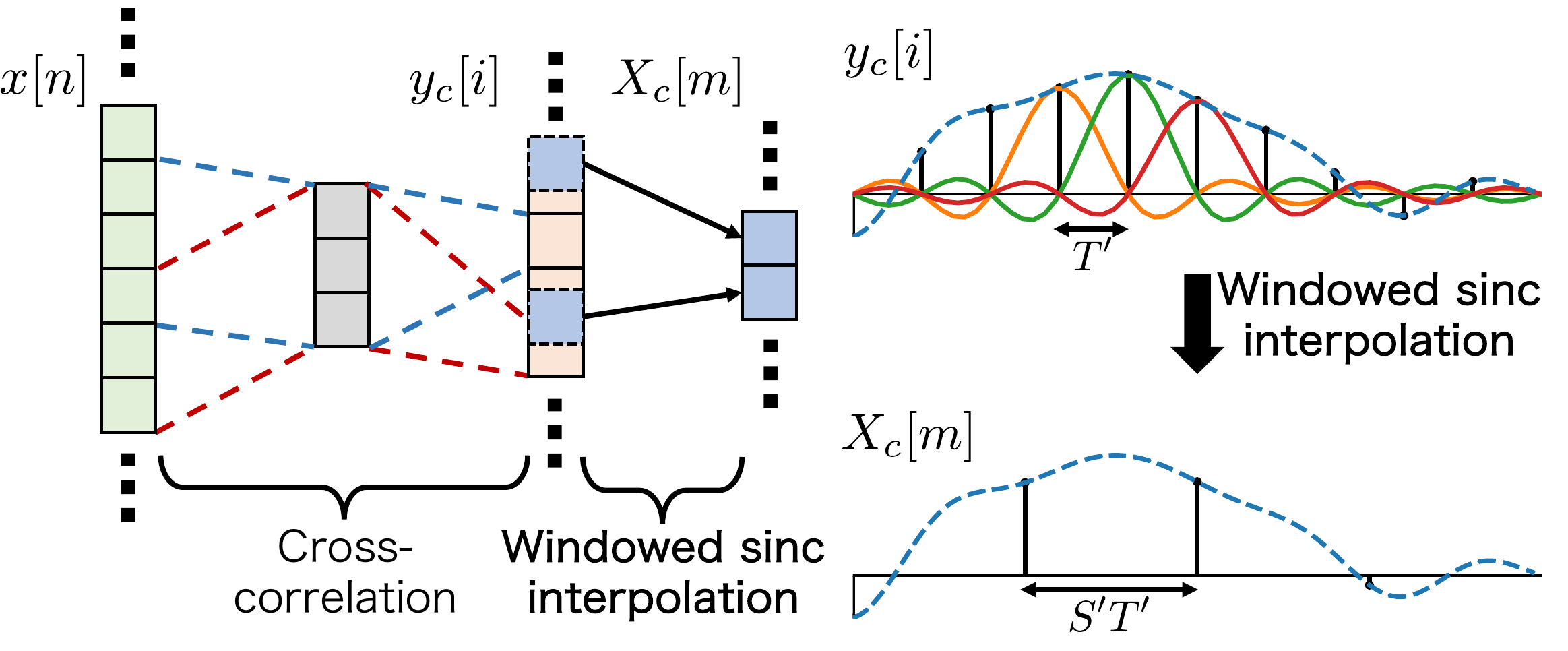}
    \subcaption{Algorithm for SFI convolutional layer}
    \label{fig:conv}
    \end{minipage}
    \hspace{0.5em}
    \begin{minipage}{0.48\hsize}
    \includegraphics[width=\linewidth]{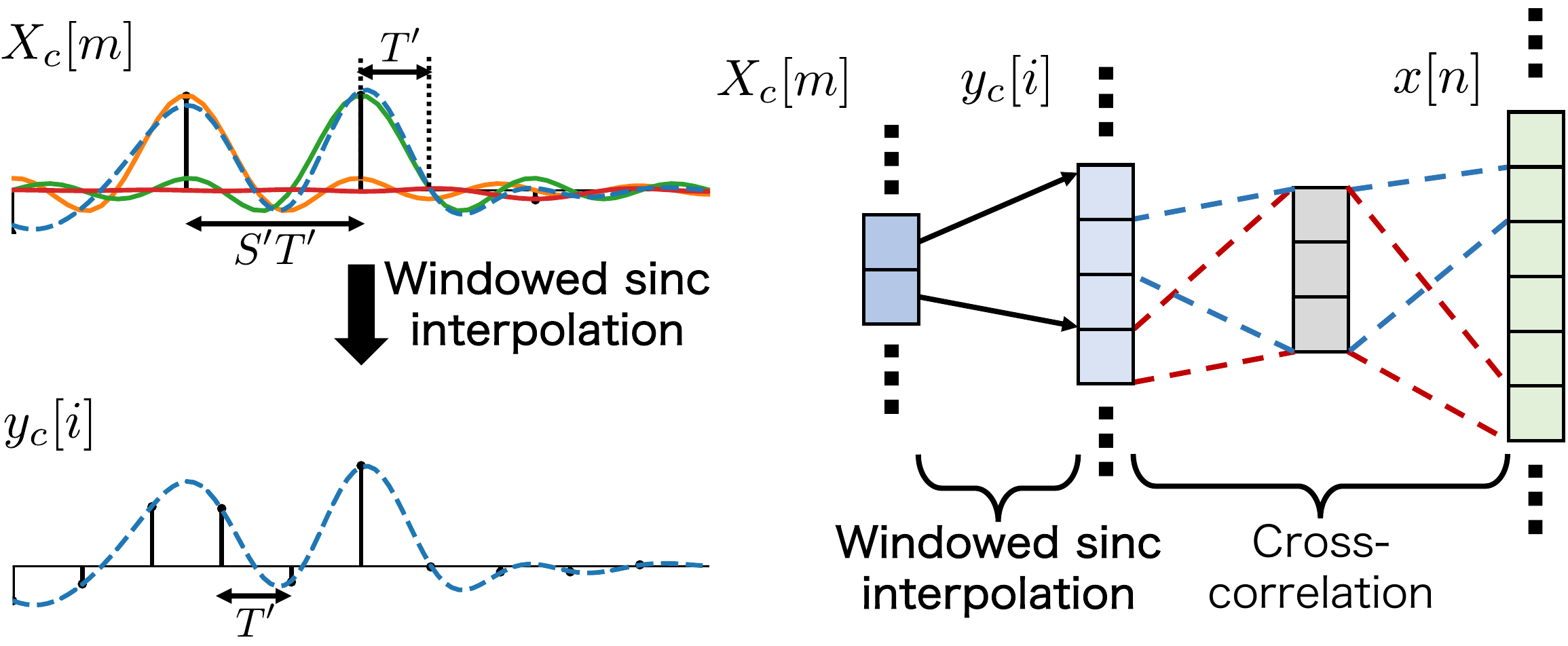}
    \subcaption{Algorithm for SFI transposed convolutional layer}
    \label{fig:transposed}
    \end{minipage} 
    \caption{Schematic illustration of proposed algorithms for SFI layers.}
    \label{fig:proposed}
\end{figure*}

\subsection{Motivation and Strategy}
To adjust $K^\txtarget$ and $S^\txtarget$ for a target SF, we previously used \eqref{eq:rounding}, as described in Section~\ref{sec:sfi-convtasnet}.
However, this adjustment method requires $K^\txtarget$ and $S^\txtarget$ to be integers.
Thus, we need another method for non-integers $K^\txtarget$ and $S^\txtarget$.

The simplest way is to round non-integers to the nearest integers.
In a preliminary experiment, we confirmed that rounding $K^\txtarget$ did not significantly affect the separation performance.
One reason for this is that, similar to a time-frequency transform, $K^\txtarget$ mainly affects spectral leakage in the frequency axis and has little impact on the time resolution of $\{X_{c}[m]\}_{c,m}$.
However, $S^\txtarget$ is not used in the weight generation process and rounding $S^\txtarget$ inevitably changes the time resolution of $\{X_{c}[m]\}_{c,m}$.
Thus, this mismatch can affect the separation performance, as we will show in Section~\ref{sec:exp}.

To resolve this mismatch, we propose algorithms of the SFI layers for non-integer $S^\txtarget$ by introducing the windowed sinc interpolation.
Fig.~\ref{fig:proposed} shows a schematic illustration of the proposed algorithms.
Since the interpolation provides a continuous-time counterpart of the discrete-time signal, it enables us to sample instants at any SF.
For simplicity, we describe algorithms for the SFI layers of SFI Conv-TasNet in the subsequent sections, but the following derivations can be extended for every pair of $\Cin$ and $\Cout$.
We also assume that $g(t)$ is an even function and $K^\txtarget$ is rounded, which we denote by $\Kround$.

\subsection{Algorithm for SFI Convolutional Layer} \label{sec:proposed_conv}
In this section, we propose a sinc-interpolation-based algorithm for the SFI convolutional layer.
The convolutional layer with a stride of $S^\txtarget$ and a padding size of $P$ first computes a cross-correlation between $\{x[n]\}_{n}$ and the weights $\{w_{c}[k]\}_{c=0,k=\lfloor-(\Kround-1)/2\rfloor}^{C-1,\lfloor(\Kround-1)/2\rfloor}$ and then decimates it with an interval of $S^\txtarget$.
The cross-correlation $y_c[i]$ is given by
\begin{equation}
    y_c[i] = \sum_{n=\lfloor-(\Kround-1)/2\rfloor}^{\lfloor(\Kround-1)/2\rfloor} x[i+n] w_c[n], \label{eq:cross-correlation}
\end{equation}
where $i$ is the discrete-time index.
The length of $y_c[i]$ is given as $I=N+2P-\Kround+1$.
Although we can decimate $y_c[i]$ with a factor of $S^\txtarget$ for an integer $S^\txtarget$, we cannot perform such decimation directly for a non-integer $S^\txtarget$.
Thus, the proposed algorithm applies the sinc interpolation to $y_c[i]$, which enables us to compute $X_c[m]$ for a non-integer $S^\txtarget$.

By invoking \eqref{eq:sinc-interp}, we can describe the interpolation of $y_c[i]$:
\begin{equation}
    \tilde{y}_c(t) = \sum_{i=\lceil t/T^\txtarget - L/2 \rceil}^{\lfloor t/T^\txtarget + L/2 \rfloor}
    y_c[i] h(t-iT^\txtarget,T^\txtarget).
    \label{eq:def_tilde_y_ct}
\end{equation}
By sampling instants from $\tilde{y}_c(t)$ with an interval of $S^\txtarget T^\txtarget$, we can obtain the pseudo time-frequency representation for non-integer $S^\txtarget$\footnote{When resampling a signal with a windowed sinc interpolation, we usually avoid aliasing by changing the second argument of $h(t,T)$. However, the decimation of a convolutional layer is not followed by any antialiasing method. Thus, we emulate this decimation in \eqref{eq:resampling_conv}.}:
\begin{equation}
    X_c[m] = \tilde{y}_c(mS^\txtarget T^\txtarget).
    \label{eq:resampling_conv}
\end{equation}
If $S^\txtarget$ is an integer, \eqref{eq:resampling_conv} reduces to decimating $y_c[i]$ by a factor of $S^\txtarget$ because $\sinc(t)=0$ for any nonzero integer $t$.

\subsection{Algorithm for SFI Transposed Convolutional Layer}
Similarly to the SFI convolutional layer, we can derive a sinc-interpolation-based algorithm for the SFI transposed convolutional layer with non-integer $S^\txtarget$.
To simplify notations, we reuse $x[n]$, $y_c[i]$, and $w_c[k]$ for the counterparts of the transposed convolutional layer.

For an integer $S^\txtarget$, a transposed convolutional layer first pads $S^\txtarget-1$ zeros between $X_c[m]$ to obtain $y_c[i]$.
It then outputs $x[n]$ as the cross-correlation between $y_c[i]$ and $w_c[i]$.

Now, we consider a continuous-time version of the transposed convolutional layer to handle a non-integer $S^\txtarget$.
The windowed sinc interpolation of $w_c[i]$ is given as
\begin{equation}
    \tilde{w}_c(t)=\sum_{i=\lfloor-(\Kround-1)/2\rfloor}^{\lfloor(\Kround-1)/2\rfloor}w_c[i]h(t-iT^\txtarget,T^\txtarget).
\end{equation}
Since $X_c[m]$ can be seen as a discrete-time signal with a sampling period of $S'T'$, i.e., $\sum_{m=0}^{M-1}X_c[m]\delta(t-mS^\txtarget T^\txtarget)$, we can write the continuous-time version of $x[n]$ as
\begin{align}
    \tilde{x}(t) &= \int_{-\infty}^{\infty}\mbox{d}\tau \sum_{m=0}^{M-1}X_c[m]\delta(t+\tau-mS'T') \tilde{w}_c(\tau) \\
    & = \sum_{i=\lfloor-(\Kround-1)/2\rfloor}^{\lfloor(\Kround-1)/2\rfloor} \tilde{X}_c(t+iT^\txtarget) w_c[i],
\end{align}
where $\delta(t)$ is Dirac's delta function and
\begin{equation}
    \tilde{X}_c(t) = \sum_{m=\lceil t/(S^\txtarget T^\txtarget) - L/(2S^\txtarget) \rceil}^{\lfloor t/(S^\txtarget T^\txtarget) + L/(2S^\txtarget) \rfloor}
    X_{c}[m] h(t- m S^\txtarget T^\txtarget, T^\txtarget). \label{eq:def_tilde_X_ct}
\end{equation}
By sampling $\tilde{x}(t)$ with an interval of $T'$, we obtain $x[n]$:
\begin{align}
y_c[i]=&\tilde{X}_c(iT^\txtarget), \label{eq:def_y_c_i}\\
x[n]=&\sum_{i=\lfloor-(\Kround-1)/2\rfloor}^{\lfloor(\Kround-1)/2\rfloor} y_c[i+n]w_c[i].
\end{align}
If $S^\txtarget$ is an integer, \eqref{eq:def_y_c_i} reduces to the zero-padding of $X_c[m]$ because $\sinc(t)=0$ for any nonzero integer $t$.

The proposed algorithms use the windowed sinc interpolation and the interpolation accuracy is determined by $L$.
Since the interpolation accuracy would affect the separation performance, we will examine their relationship in Section~\ref{sec:exp_compt}.

\section{Experiments} \label{sec:exp}
\subsection{Experimental Setup}
To evaluate the effectiveness of the proposed algorithms, we conducted music source separation experiments using the MUSDB18-HQ dataset \cite{MUSDB18HQ}.
This dataset has 150 tracks, each of which consists of four instruments (vocals, bass, drums, and other).
We used the official data split of 86, 14, and 50 tracks for training, validation, and test, respectively.
As an evaluation measure, we used the signal-to-distortion ratio (SDR) obtained with the BSSEval v4 toolkit\cite{bsse}.
To reduce the dependency on initialization, we trained models using four random seeds and computed the averages and standard errors of the SDRs.

We used the same training setup as in \cite{sfi-convtasnet}.
As a source separation model, we chose SFI Conv-TasNet using the frequency-domain filter design, which achieved the highest separation performance in \cite{sfi-convtasnet}.
The latent analog filter was a modulated Gaussian filter. Its impulse response is given by
\begin{equation}
    f(t) = 2 \sqrt{2\sigma^2\pi} \exp\left(-\frac{\sigma^2t^2}{2}\right) \cos(\mu t + \phi),
    \label{mgf}
\end{equation}
where $\mu$ is the center angular frequency, $\sigma$ is the parameter corresponding to the bandwidth of the filter, and $\phi$ is the initial phase.
These parameters were initialized as in \cite{sfi-convtasnet} and were trained jointly with other DNN parameters.
The SFI Conv-TasNet was trained with a batch size of 12 for 250 epochs by RAdam\cite{radam} wrapped in a LookAhead optimizer \cite{lookahead}.
The same data augmentations used in \cite{sfi-convtasnet} were applied.
The loss function was the negative scale-invariant source-to-noise ratio (SI-SNR) between the estimated and groundtruth signals.
Since the SDR is scale-dependent, we used the scaling method presented in \cite{meta-learning}: $\{\alpha_i\}_{j=1}^{J} = \argmin_{\{\alpha_j\}_{j=1}^{J}} \sum_{n=0}^{N-1}(x[n] - \sum_{j=1}^{J} \alpha_j \hat{s}_j[n])^2$, where $\alpha_j$ is the scale for source $j$.

We compared the following four methods of handling non-integer strides.
\emph{Rounding} is a method to round $S^\txtarget$ to the nearest integer.
\emph{Resampling-near} avoids non-integer $S^\txtarget$ by resampling a mixture signal at the nearest SF where the corresponding stride becomes an integer, separating the resampled signal with the trained model, and resampling the separated signals back to the original SF.
\emph{Resampling-trained} resamples a mixture signal at the trained SF.
The signal resampling was implemented with \texttt{librosa}\cite{librosa} as in \cite{sfi-convtasnet}.
\emph{Proposed} uses the proposed algorithms for handling non-integer $S^\txtarget$.
The windowed sinc interpolation was implemented with \texttt{torchaudio}\cite{torchaudio}.
For $g(t)$, we used the Kaiser window with the default parameters of \texttt{torchaudio.functional.resample}.
All methods used the same trained models.
Following the experimental conditions used in\cite{sfi-convtasnet}, the SFs of the training and validation data were set to 32~\si{\kilo\hertz} and the kernel size and stride were set to 5 and 2.5 ms ($K^\txtrain=160$ and $S^\txtrain=80$), respectively.
The methods were evaluated at an SF of 11.025, 22.05, and 44.1~\si{\kilo\hertz}, where $S^\txtarget$ becomes $27.5625$, $55.125$, and $110.25$, respectively.
In addition to these popular SFs, we evaluated the methods with an in-between SF, 16.538~\si{\kilo\hertz}, to show the generality of the proposed method.

\subsection{Relationship Between Interpolation Accuracy and Separation Performance} \label{sec:exp_compt}
\begin{table}[t]
    \centering
    {
    \caption{SDRs [\si{\decibel}] of Proposed with varying $L$ for vocals}
    \begin{tabular}{ccccc}
        \toprule
       \multirow{2}{*}{$L$} & \multicolumn{4}{c}{SF [\si{\kilo\hertz}]} \\
        & 11.025 & 16.538 & 22.05 & 44.1 \\
        \midrule
         2 & 2.3 $\pm$ 0.3 & 1.5 $\pm$ 0.3 & 2.4 $\pm$ 0.1 & 1.8 $\pm$ 0.1 \\
         4 & 4.7 $\pm$ 0.2 & 4.8 $\pm$ 0.0 & 5.4 $\pm$ 0.1 & 5.5 $\pm$ 0.1 \\
         8 & 4.6 $\pm$ 0.2 & 5.2 $\pm$ 0.1 & 5.6 $\pm$ 0.0 & 5.8 $\pm$ 0.1 \\
         16 & 4.7 $\pm$ 0.2 & 5.3 $\pm$ 0.1 & 5.6 $\pm$ 0.1 & 5.8 $\pm$ 0.1 \\
         32 & 4.7 $\pm$ 0.2 & 5.4 $\pm$ 0.1 & 5.6 $\pm$ 0.1 & 5.8 $\pm$ 0.1 \\
         64 & 4.7 $\pm$ 0.2 & 5.4 $\pm$ 0.1 & 5.6 $\pm$ 0.1 & 5.8 $\pm$ 0.1 \\
        \bottomrule
    \end{tabular}
    \label{tab:sdr_varing_l}
    }
\end{table}

We first examined the relationship in Proposed between separation performance and interpolation accuracy.
Table~\ref{tab:sdr_varing_l} lists the averages and standard errors of SDRs with varying $L$ for vocals.
The SDRs approximately increased as $L$ increased, showing that lowering the interpolation accuracy degrades the separation performance.
At all SFs, the SDRs were saturated for $L\geq 16$ and this tendency was observed in the results of bass, drums, and other.
Hence, we decided to use $L=16$.

Since $L$ also involves the processing time, we measured the processing times for a $60$-s signal using an NVIDIA GeForce RTX 3090 GPU.
At an SF of $11.025$~\si{\kilo\hertz}, the processing time averaged over ten trials was 72.8 ms.
Although this is longer than Rounding (52.1 ms), it is fast enough in practice and can be reduced by brushing up the implementation of Proposed.

\subsection{Comparison with Existing Methods}
\label{sec:separation}
\begin{figure*}[t]
    \centering
    \begin{minipage}{1\hsize}
    \centering
    \includegraphics[width=0.68\linewidth]{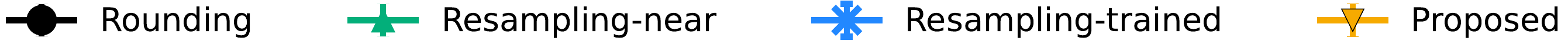}
    \end{minipage}
    \\
    \vspace{0.3em}
    \begin{minipage}{0.24\hsize}
        \centering
        \includegraphics[width=\linewidth]{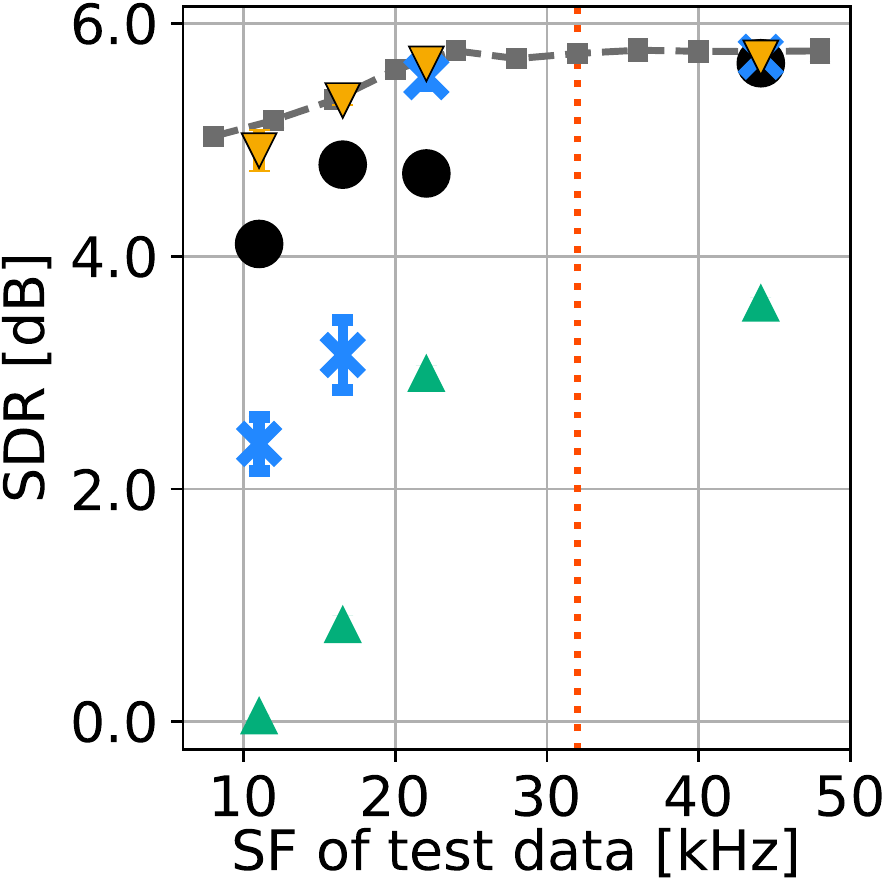}
        \vspace{-1.3\baselineskip}
        \subcaption{vocals}
    \end{minipage}
    \begin{minipage}{0.24\hsize}
        \centering
        \includegraphics[width=\linewidth]{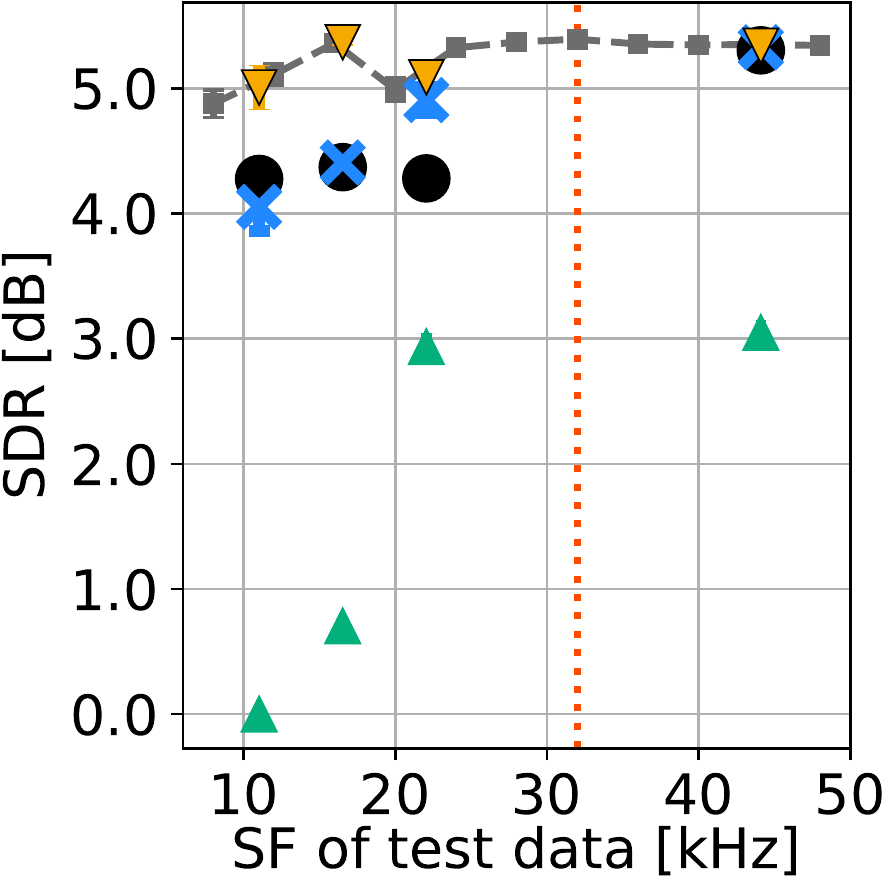}
        \vspace{-1.3\baselineskip}
        \subcaption{bass}
    \end{minipage}
    \begin{minipage}{0.24\hsize}
        \centering
        \includegraphics[width=\linewidth]{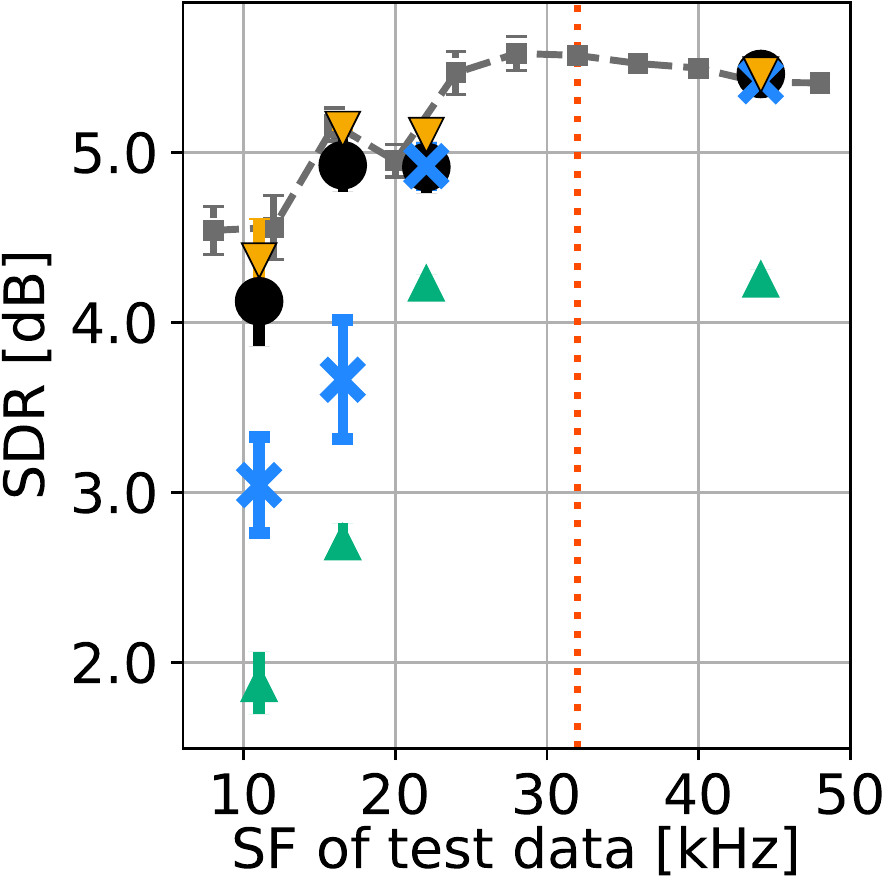}
        \vspace{-1.3\baselineskip}
        \subcaption{drums}
    \end{minipage}
    \begin{minipage}{0.24\hsize}
        \centering
        \includegraphics[width=\linewidth]{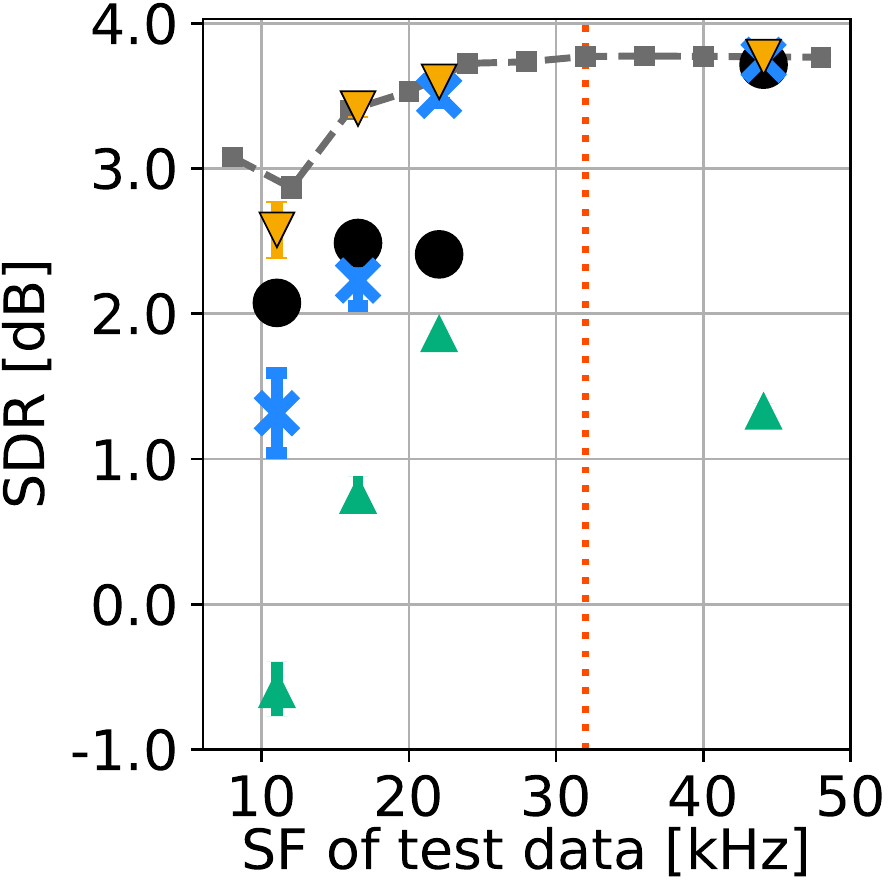}
        \vspace{-1.3\baselineskip}
        \subcaption{other}
    \end{minipage}
    \caption{
        SDRs of Rounding, Resampling-near, Resampling-trained, and Proposed. 
        Error bars represent standard errors and red dotted line denotes the trained SF.
        For reference, gray dashed line shows SDRs at SFs where $K^\txtarget$ and $S^\txtarget$ are integers.
    }
    \label{fig:sdr}
\end{figure*}

Fig.~\ref{fig:sdr} shows the averages and standard errors of SDRs for all methods per instrument.
As a reference, we also show SDRs at SFs where $K^\txtarget$ and $S^\txtarget$ are integers (gray dashed curve), which we call \emph{Reference}.
SDRs of Resampling-near significantly decreased compared with Reference.
Although the other methods had SDRs consistent with Reference at an SF of 44.1~\si{\kilo\hertz}, Rounding and Resampling-trained quickly decreased as the SF decreased.
At the lower SFs, SDRs of Rounding were around one decibel below Reference for vocals, bass, and other.
One reason for the performance drop is that the change in time resolution caused by Rounding increases as the SF decreases.
This result shows that rounding $S^\txtarget$ degrades the separation performance at lower SFs.
Resampling-trained had higher SDRs than Rounding at an SF of 22.05~\si{\kilo\hertz}.
However, its SDRs dropped significantly as the SF decreases, which is consistent with the results observed in \cite{sfi-convtasnet}.
By contrast, Proposed provided more consistent SDRs with Reference at all SFs for all instruments.
These results clearly demonstrate the effectiveness of the proposed algorithms.

The SDR gaps between Proposed and Rounding for vocals, bass, and other were greater than those for drums.
The change in time resolution caused by rounding $S^\txtarget$ corresponds to that in the SF of an input signal for the mask predictors.
Thus, it should strongly affect the separation performance for the audio signals containing pitched sounds.
The drums are unpitched instruments and the change in time resolution can be less significant compared with pitched sounds.
This observation leads to a conjecture that slightly changing the time resolution of an input is an adversarial attack for the separation of pitched sounds.
Its investigation is beyond the scope of this paper and we leave it as future work.

\section{Conclusion}
We proposed algorithms of the SFI layers for non-integer strides.
The proposed algorithms use the windowed sinc interpolation to bridge two signals of different SFs.
For the SFI convolutional layer, the interpolation is applied to the cross-correlation between an input and the weights, which enables us to sample instants with a desired SF.
For the SFI transposed convolutional layer, the interpolation is applied to an input.
By using these algorithms, we can handle non-integer strides in SFI Conv-TasNet.
Experimental results showed that the proposed method outperforms the rounding- and signal-resampling-based methods at lower SFs.

\bibliographystyle{IEEEtran}
\bibliography{abbrev, reference}
\end{document}